\documentclass[12pt,preprint]{aastex}
\begin{document}

\title{The Highly Eccentric Pre$-$Main Sequence Spectroscopic Binary RX~J0529.3$+$1210}

\author{G. N. Mace\altaffilmark{1,2}, L. Prato\altaffilmark{1},
  L.H. Wasserman\altaffilmark{1}, G. H. Schaefer\altaffilmark{3},
O.G. Franz\altaffilmark{1}, and M. Simon\altaffilmark{4}}

\altaffiltext{1}{Lowell Observatory, 1400 West Mars Hill Road,
Flagstaff, AZ 86001; gmace@lowell.edu}

\altaffiltext{2}{Department of Physics and Astronomy, Northern
Arizona University, Flagstaff, AZ 86011}

\altaffiltext{3}{The CHARA Array of Georgia State University, Mount Wilson
  Observatory, Mount Wilson, CA 91023}

\altaffiltext{4}{Department of Physics and Astronomy, State University of New
York, Stony Brook, NY 11794-3800}

\begin{abstract}

The young system RX~J0529.3$+$1210 was initially identified as a single$-$lined
spectroscopic binary. Using high$-$resolution infrared spectra,
acquired with NIRSPEC on Keck II, we measured radial velocities for the
secondary. The method of using the infrared regime to convert
single$-$lined spectra into double$-$lined spectra, and derive the mass ratio
for the binary system, has been successfully used for a number of young,
low-mass binaries. For RX~J0529.3$+$1210, a long-period (462
days) and highly eccentric (0.88) binary system, we determine
the mass ratio to be 0.78 $\pm$ 0.05 using the infrared double-lined
velocity data alone, and
0.73 $\pm$ 0.23 combining visible light and infrared data in a full
orbital solution. The large uncertainty in the latter is the result
of the sparse sampling in the infrared and the high eccentricity: the stars
do not have a large velocity separation during most of their $\sim$1.3 year
orbit.  A mass ratio close to unity, consistent with the high end of
the 1$\sigma$ uncertainty for this mass ratio value, is inconsistent
with the lack of a visible light detection of the secondary
component.  We outline several scenarios for a color difference in the
two stars, such as one heavily spotted component, 
higher order multiplicity, or a unique evolutionary stage, favoring detection
of only the primary star in visible light, even in a mass ratio $\sim$1
system.  However, the evidence points to a lower ratio.  Although RX~J0529.3$+$1210
exhibits no excess at near-infrared wavelengths, a small 24~$\mu$m excess
is detected, consistent with circumbinary dust.  The properties of this
binary and its membership in $\lambda$ Ori versus a new nearby stellar
moving group at $\sim$90~pc are discussed. We speculate on the origin of
this unusual system and on the impact of such high eccentricity, the largest
observed in a pre$-$main sequence double$-$lined system to date, on the
potential for planet formation.

\end{abstract}


\section{Introduction}

Understanding the process of star formation requires
reliable observations of fundamental stellar properties so that theoretical models can be tested.  The copious population of young binary systems, however, complicates the problem.  The binary fraction can be high \citep{2000prpl.conf..703M}, and may depend on the density of the star forming region \citep{2003ApJ...583..358B} and spectral type of the primary \citep{2006ApJ...640L..63L}, underscoring the importance of their study. Fortuitously, binary stars serve two purposes.  One, characterization of their frequency, separation distribution, and mass ratio distribution for a given
star forming region (SFR) provides clues to the broad star forming
properties (angular momentum, density, turbulence, etc.) of the parent
molecular cloud.  Two, the special class of very small separation 
spectroscopic binaries with periods sufficiently short to enable the measurement of the individual
stellar velocities, and thus the system's mass ratio, are potential targets
for the dynamical determination of individual component stellar masses (e.g., Steffen et al. 2001; Prato et al. 2002a; Boden et al. 2005; Stassun et al. 2007). Knowledge of absolute masses and observable properties, such as effective temperature and luminosity, plays a key
role in the improvement of pre-main sequence (PMS) evolutionary models \citep{2001ApJ...553..299P}.

Once spectroscopic binaries have been identified, it is necessary to
characterize their properties.  It is often the case with low mass-ratio
systems that they are identified as single$-$lined spectroscopic binaries when observed
in visible light (Mazeh et al. 2002), in which case the large
difference in flux between the primary and secondary at short wavelengths
prevents detection of the secondary component and thus the measurement of
the mass ratio for the system.  In the Raleigh-Jeans regime, however, flux scales much less 
steeply as a function of mass. Thus, by observing single-lined spectroscopic binaries
with infrared (IR) spectroscopy we are able to improve our chances of
detecting the lower-mass secondary not seen in visible light. This technique was
initially outlined in Prato et al. (2002b) and Mazeh et al. (2002, 2003). Our
primary motivation for observing RX~J0529.3$+$1210 was to use this IR approach to
determine the mass ratio of the system by converting it into a double-lined
spectroscopic binary. Given that the secondary had not been detected in visible light,
we anticipated a relatively low mass ratio for RX~J0529.3$+$1210.

With the advent of the {\it R\"ontgensatellit (ROSAT)} all-sky
survey, a number of researchers began a search for X-ray sources with
optical counterparts of brightness consistent with membership in nearby
SFRs, motivated in part by the goal of detecting the ``post-T Tauri''
population postulated by Herbig (1978).  In follow up observations of X-ray sources near Taurus, RX~J0529.3$+$1210 was identified variously as a PMS star (Neuhauser et al. 1997;
Magazz\`u et al. 1997) and a post-T Tauri star (Magazz\`u et al. 1999); 
high-resolution visible light spectroscopy revealed its spectroscopic binary nature
\citep{neu97}.  \citet{tor02} undertook a seven year campaign to characterize
the orbits of all spectroscopic binaries identified in the X-ray sample
of \citet{neu95} and \citet{neu97}, including RX~J0529.3$+$1210.  Table 1 summarizes
the general properties of this system.  The $\sim$3 dozen high-resolution
visible light spectra taken
of this binary suggested the presence of a secondary star; however, a conclusive
identification was not possible.  A single-lined orbital solution was determined, although,
probably owing to the extremely high eccentricity of the system the
uncertainties are relatively large.  

This paper describes the results of using high resolution IR spectroscopy to
observe RX~J0529.3$+$1210 and determine the component radial
velocities. RX~J0529.3$+$1210 is the most eccentric pre$-$main sequence spectroscopic binary known to date. This provides a unique context in which to speculate
on the formation of the system and on the impact of the stellar dynamics on potential planet
formation. In \S2 we briefly describe our observations and data reduction.
Our analysis and results appear in \S3. Section 4 provides a discussion,
and \S5 summarizes our findings.

\section{Observations and Data Reduction}

Our observations were made during six epochs between 2002 January and 2004
December with the Keck~II 10-m telescope on Mauna Kea. The UT dates of
observation are listed in Table 2.
$H$-band data, at a central wavelength of $\sim$1.555 $\mu$m, were
obtained using the facility near-infrared, cross-dispersed, cryogenic
spectrograph NIRSPEC (McLean et al. 1998; 2000). NIRSPEC employs a 1024
$\times$ 1024 ALADDIN InSb array detector. We used the 0.288$''$ (2 pixel)
$\times$ 24$''$ slit, yielding an OH night sky emission line determined resolution of R$=$25,000. Source acquisition was accomplished with the slit viewing camera, SCAM, which utilizes a 256 $\times$ 256 HgCdTe detector with 0.18$''$ pixels. RX~J0529.3$+$1210 has a 2MASS $H$-band magnitude of 9.40; integration times for individual frames were between 60 and 240~s. We nodded the telescope
10$''$ between two positions on the slit to allow for background subtraction between sequential spectra.

All spectroscopic reductions were made with the REDSPEC package, software written at
UCLA by S. Kim, L. Prato, and I. McLean specifically for the analysis of
NIRSPEC data\footnote{See: http://www2.keck.hawaii.edu/inst/nirspec/redspec/index.html},
following the procedures outlined by Prato et al. (2002a). We used the central
order, 49, which covers 1.545 to 1.567 $\mu$m at our setting; this order has three advantages.  One, it is rich in both atomic and molecular lines and is therefore suitable for identifying
spectra of both warm and cool stars.  Two, the OH night sky emission lines
across order 49 are numerous and well-distributed, yielding accurate and precise dispersion
solutions. Three, this order has the advantage of lacking prominent telluric absorption lines. Consequently, we did not have to divide by telluric standard star spectra. 

On UT dates 2004 December 24 and 2008 January 17 images were taken of
RX~J0529.3$+$1210, point spread function (PSF) stars,
and photometric standards at the Keck II telescope
with the NIRC2 camera behind the adaptive optics (AO) system \citep{2000SPIE.4007....2W}.
The plate scale was 0.01$''$ per pixel and integration times were 0.18 s for
the L' filter, 0.20 s for the K' filter, and 1.0 s for the narrow K-continuum
filter.  Ten coadds were made per image,
and two images were obtained at each location in a five-point box dither pattern.  Seeing conditions were excellent for both nights.
With the use of customized IDL reduction and analysis routines,
images were flat-fielded, cleaned of bad pixels, and
compared with PSF stars to search for higher order multiplicity
and evidence for extended sources.  L' and K-continuum filter data from
2008 January 17 (UT) were reduced with standard photometric techniques,
including flat-fielding, background subtraction, aperture photometry, and
airmass correction.

\section{Analysis and Results}

The spectra from the six epochs of observation are shown in Figure 1.
Individual stellar radial velocities were measured by using two-dimensional cross
correlation (e. g., Zucker \& Mazeh 1994); observed or synthetic
spectral templates were shifted in relative velocity to maximize the correlation with the observed binary spectrum,  thus identifying the component radial velocities.  The observed template spectra \citep{pra02a} that best matched our data were BS 8086 and GL 436, spectral types K7 and M2.5, consistent with \citet{tor02}.  The corresponding best rotational velocities were 20 km s$^{-1}$ for the primary and 25 km s$^{-1}$ for the secondary.  Flux ratios determined by cross-correlation range from 0.50 to 0.69 and have an average value of $\sim$0.6. 

\citet{2008AJ....135.1659S} employ synthetic spectra in their final radial velocity
analysis of the components of the young spectroscopic binary, Haro 1-14c.  In
addition to using observed templates, we also fit the components of
RX~J0529.3$+$1210 with synthetic template spectra, calculated from the
updated NextGen models \citep{1999ApJ...512..377H}.  This is desirable because
observed template spectra have inherent radial velocity uncertainties, whereas
the synthetic template spectra have no
associated velocity measurement uncertainties.  Thus, by using synthetic templates, the
uncertainties in the component radial velocities of a spectroscopic binary
are likely attributable to factors such as the signal to noise ratio of the
observed spectroscopic binary spectra and imperfect matches to model stellar
atmosphere spectra.  For RX~J0529.3$+$1210 we were unable to determine consistent
velocity solutions using synthetic templates for both the
primary and secondary stars because of the poor fit of a synthetic secondary to the
observed spectra.  However, the combination of a synthetic spectrum
for the primary component (rotated to 30 km s$^{-1}$) and the GL~436 observed
template for the secondary (rotated to 25$-$30 km s$^{-1}$) was successful.
The velocities derived from this approach appear
in Table 2, columns (3) and (4).  The velocity uncertainties associated with the
primary star are 0.5~km s$^{-1}$ and with the secondary star
are 2.0~km s$^{-1}$.  Analysis of the spectra with only the observed templates yields results
that are consistent to within 1$\sigma$ with the mixed approach presented here.

Following Wilson (1941), we plot the six epochs of the
primary versus secondary radial velocities (Figure 2), extracted from our IR
spectra, and derive a mass ratio of q$=$0.78$\pm$0.05 and a center-of-mass
velocity of $\gamma=+$19.01 $\pm$ 0.87~km~s$^{-1}$.  We combined our radial
velocities for the primary and secondary stars with the 34 measurements
for the primary star presented in \citet{tor02}
and solved for the best-fit orbital elements in the $\chi^2$ sense. The
orbital fit is a standard least-squares program using the Levenberg-Marquardt
method taken from \citet{1992nrfa.book.....P}.  Initial guesses for the
solution are found by using an amoeba search routine also  from
\citet{1992nrfa.book.....P}.  The orbital elements are given in Table 3.  The
mass ratio and the center-of-mass velocity found from the single- and
double-lined velocities together are consistent with those derived from
Figure 2 to within 1$\sigma$, although the mass ratio calculated from the
full orbital solution, 0.73$\pm$0.23, has a large associated
uncertainty. Because the
primary star velocity measurements dominate the data set, the orbital
parameters are similar to those found in \citet{tor02}.  Determination
of the orbital solution from only the double-lined velocities yields
results consistent with the combined single- and double-lined solution
but with much larger uncertainties.  We describe future work to improve
our orbital fit in \S4.4.

Figure 3 shows the primary star velocities from \citet{tor02} along with IR
determined radial velocities for both the primary and secondary stars
plotted as a function of phase. The phases for the IR observations were
computed by taking the difference in the heliocentric Julian Day (HJD) from
the last observation made by Torres et al. (2002) and the HJDs of our
observations and dividing by our period (Table 3).  The errors associated with the
velocities are typically a few km~s$^{-1}$ for the \citet{tor02} data,
0.5 km~s$^{-1}$ for our primary star data, and 2.0 km~s$^{-1}$ for our
secondary data.  Our orbital solution is also plotted.  Clearly there is
scatter of up to several sigma in many of the data points, particularly among the
visible light data with velocities closer to the center-of-mass velocity of the system,
compared to the orbital solution.  \citet{tor02} give a typical radial velocity uncertainty
for all their targets of 0.5~km~s$^{-1}$, probably an underestimate in the case of
RX~J0529.3$+$1210.  The IR data are within 1$-$2~$\sigma$ of the orbital solution.
The sample of PMS spectroscopic binary star eccentricities as a function of period,
based on data from \citet{mel01}, is plotted in Figure 4 and indicates that
RX~J0529.3$+$1210 is the most eccentric PMS spectroscopic binary known
to date.  The combination of the system orientation and high eccentricity
results in fairly low velocity amplitudes during the majority of the
orbital period.  Until the high-velocity cusp of the radial
velocity versus phase curve is well-sampled, the
uncertainty in the elements of this system will remain large.

Near-IR J$-$H and H$-$K colors for RX~J0529.3$+$1210 were calculated
from 2MASS JHK magnitudes and plotted on a color-color diagram.  The
location in the color-color plane of this source is consistent with that of a
late K or early M dwarf.  There is no near-IR excess evident.  From our
Keck$+$NIRC2 data we determine a K-continuum magnitude of 9.23$\pm$0.27,
consistent with 2MASS measurements (Table 1), and an L'-band magnitude
of 9.05$\pm$0.11 from our 2008 Keck AO images.  The resultant K$-$L' color
is thus 0.18$\pm$0.29 magnitudes.  

In the AO images from 2008 January 17, RX~J0529.3$+$1210 appears to be extended in comparison with the observed single star PSFs.  Modeling RX~J0529.3$+$1210 as a binary with the single star DN Tau used as a PSF yields a separation of $0.018''\pm0.006''$ at a position angle of $223^\circ\pm18^\circ$ and a secondary-to-primary K-band flux ratio of $0.66\pm0.18$.  The reliability of the solution is likely to be low because of the difficulty in measuring separations this far below the diffraction limit.  We also note the possibility that RX~J0529.3$+$1210 could be broadened by a lower AO correction rate during the observations as compared with that of the single star.  However, the consistency of the flux ratio with the H-band spectroscopic values determined by cross-correlation provides support that we are seeing evidence of the companion in RX~J0529.3$+$1210.  We do not see any extension in the 2004 December 24 images of RX~J0529.3$+$1210, which is consistent with the binary being near periastron, whereas in 2008 January the companion was approaching apastron.

\section{Discussion}

\subsection{Where and How Old is RX~J0529.3$+$1210?}

The projected location of RX~J0529.3$+$1210 is associated with a high-density
area in the
CO maps of \citet{dol01} for the $\lambda$ Ori region.  However, \citet{dol01}
find a mean radial velocity for the strong lithium sources identified in
$\lambda$ Ori of 24.5~km~s$^{-1}$, with a dispersion of only 2.3~km~s$^{-1}$.
Our center-of-mass velocity for RX~J0529.3$+$1210 is 18.38~km~s$^{-1}\pm0.30$,
indistinguishable from that found by \citet{tor02}.  Thus,
on the basis of radial velocities alone, it seems unlikely
that the system is associated with $\lambda$ Ori since RX~J0529.3$+$1210's
center-of-mass velocity is inconsistent with that of $\lambda$ Ori at the 3$\sigma$ level.

Using the 2MASS JHK magnitudes of RX~J0529.3$+$1210, an effective temperature
(T$_{eff}$) for a K7/M0 of 3900~K \citep{2003ApJ...593.1093L,bro06}, and the nominal distance to $\lambda$ Ori (400~pc; Barrado y Navascu\'es 2005) we have estimated the luminosity and placed the system on an H-R diagram.  We find L $=$ 3.44 $\pm$ 0.01 L$_{\odot}$, yielding an age of $\sim$0.1 Myr using the PMS tracks of \citet{pal99}.  Accounting for a companion star with luminosity equal to
that of the primary results in an age of $\sim$0.5~Myr.  Using the tracks of \citet{bar98}, or absolute K magnitude and the tracks of \citet{2000A&A...358..593S}, we obtain similar age estimates.

\citet{bar05} estimates the age of $\lambda$ Ori to be between
3 and 10 Myr, \citet{bar07} find an age of 5~Myr, and \citet{dol01} describe a
spread in ages from 1 to 10 Myr.
A $<<$1~Myr object is expected to be at least somewhat embedded in its natal
cloud and associated with circumstellar material,
yet near-IR colors from 2MASS magnitudes
indicate zero extinction, no veiling is detected in the spectra, and the
H$\alpha$ emission line equivalent width (Table 1) is only 2\AA.  We
consider two possible mechanisms for disk dissipation in this system,
photoevaporation from nearby hot stars and the orbital dynamics of
RX~J0529.3$+$1210 itself.

The projected separation of the
B8 star HD 36104 \citep{dol01} from RX~J0529.3$+$1210 is $\sim$1~pc if
both are assumed to be at a distance of 400~pc. \citet{dol01}
discuss the surprising lack of evidence for accretion disks around the PMS
stars in the central $\lambda$ Ori cluster and suggest that photoevaporation
and/or possibly a supernova event 1$-$2 Myr ago played a role in the dispersion
of disks in the local young low-mass stellar population.  RX~J0529.3$+$1210,
located just north of the central cluster in the Barnard 30 dark cloud, could
have experienced photoevaporation from the nearby B star at a young age,
obliterating circumstellar material, although this is probably unlikely given the long survival time of proplyds in the Trapezium \citep{1998AJ....115..263O}.

Given the extremely high eccentricity of this binary, however, the
action of the companion star may have precluded formation of, or
dissipated any, circumstellar
material.  With an assumed primary star mass of 0.75 M$_{\odot}$
and the mass function from Torres et al. (2002), we find
a minimum secondary star mass of 0.40 M$_{\odot}$, and
thus a minimum total mass of 1.15 M$_{\odot}$.  In conjunction
with the 461.89 day period, this yields a periastron separation
of 0.15~AU and an apastron separation of 2.30~AU. It is unknown if orbital evolution may have occurred, or even
whether it is possible that this unusual system formed by capture,
but if RX~J0529.3$+$1210 is located in $\lambda$ Ori with an age of
$<$1~Myr, then it seems likely that the system formed in a similar
configuration to its present one.

Independent of the potentially harsh $\lambda$ Ori environment and the
orbital dynamics of RX~J0529.3$+$1210, the system manifests spectra
of similar surface gravity to dwarf stars, as well as a relatively
small lithium equivalent width, in comparison to classical T Tauri stars \citep{1989AJ.....98.1444S,2005fost.book.....S}.  The lithium and H$\alpha$ equivalent widths of \citet{dol01} for $\lambda$ Ori average greater than twice what is found for RX~J0529.3$+$1210 and are consistent with the classical and weak-lined T-Tauri limits defined by \citet{1998AJ....115..351M}.   According to these limits, RX~J0529.3$+$1210 would be a post-T Tauri star and have one of the lowest lithium and H$\alpha$ equivalent widths in the $\lambda$ Ori region.  These characteristics are inconsistent
with an age of $<$1~Myr.  

Alternatively, RX~J0529.3$+$1210 could be an older, closer object.
\citet{mam07} describes a new candidate moving group, 32 Ori, consisting of
a small cluster of X-ray bright, late type stars.  The 
$\sim$10 young stars identified in the group are located
around 5$^h$ 20$^m$ to 5$^h$ 30$^m$ and $+6\deg$, at
the proposed distance of $\sim$90~pc.  RX~J0529.3$+$1210 is about 9~pc from the central
clump of objects in 32 Ori and is one of the members
used by Mamajek to define the common proper motion group (E. Mamajek 2008,
private communication).  Given the proper motion of RX~J0529.3$+$1210,
pmRA $=$  4.1 $\pm$ 5.8 mas/yr and pmDec $=$ $-$30.7 $\pm$ 5.8 mas/yr,
compared to the proper motions of the stars $\lambda$ Ori (pmRA $=$ 0.8 $\pm$ 1.5,
pmDec $=$ $-$2.3 $\pm$ 1.5) and 32 Ori (pmRA $=$ 6.57 $\pm$ 1.15,
pmDec $=$ $-$32.45 $\pm$ 0.48), evidence in support of membership in the 32~Ori
group is compelling \citep{1997A&A...323L..49P, 2004AJ....127.3043Z}.

Again combining the 2MASS JHK magnitudes of RX~J0529.3$+$1210,
an effective temperature of 3900~K, and a distance now of 90~pc, we find L $=$ 0.17 $\pm$ 0.02 L$_{\odot}$, giving an age for RX~J0529.3$+$1210 on the tracks of \citet{pal99} of $\sim$15$\pm$5~Myr,
consistent with the 25$\pm$10~Myr age derived by \citet{mam07} for the
candidate group members.  \citet{mam07} lists a group radial velocity of 18~km~s$^{-1}$, in excellent agreement with our measured center-of-mass velocity (Table 2).

Morales Calderon (2008, in prep) has observed the $\lambda$ Ori region with
the {\it Spitzer} space telescope and finds a 3~$\sigma$ detection of RX~J0529.3$+$1210
at 24~$\mu$m with a 20~\% excess above a 3900~K photosphere.  For a star
with L$=$0.17~L$_{\odot}$, the equilibrium temperature of a black-body
grain with peak emission at 24~$\mu$m corresponds to a distance of 4.36~AU.
A more realistic treatment of the dust grain distribution would necessarily
take into account the additional flux from the secondary star, yielding a
larger distance from the center-of-mass of the system to the putative dust.
However, given the estimated apastron separation of 2.30~AU, a lower limit
of 4.36~AU for the dust radius from the system center illustrates the plausibility of a 
circumbinary debris disk.  Additional {\it Spitzer} observations at
shorter wavelengths have been taken and should reveal more information
regarding the extent and location of the dust (Mamajek 2008, in preparation).

The lack of J$-$H, H$-$K, and K$-$L excesses is hardly surprising.  For
black-body grains with a peak wavelength in the L band, the corresponding
disk temperature occurs at a distance of $\sim$0.1~AU, nearly coincident with the
binary periastron.  Stable dust in this system is most likely to
be located in a circumbinary distribution.  Furthermore, if the closer
distance and therefore older age for this system implied by
membership in 32 Ori are correct, then
the presence of an evolved debris disk would not be unusual \citep[e.g.,][]{tri08}.

Finally, the fact that the RX~J0529.3$+$1210 secondary was possibly detected in our AO
images strongly supports the $\sim$90~pc distance.  For the projected separation
of 0.018", determined using PSF fitting, and a distance of 90~pc, the corresponding
distance in AU is 1.6$\pm$0.54, not too different from
our estimated apastron of 2.30~AU.  For
a distance of 400~pc the separation would be 7.2$\pm$2.4 AU.
Based on the currently available data, it therefore appears that RX~J0529.3$+$1210
is associated with the new 32 Ori group, not with $\lambda$ Ori, and
has an age of $\sim$15~Myr.

\subsection{The Mass and Flux Ratios of RX~J0529.3$+$1210}

\citet{tor02} note that the appearance of the velocity correlation function for
RX~J0529.3$+$1210 suggests the presence of at least one other star in the system.
For a primary star mass of 0.75~M$_{\odot}$, the mass function implies a minimum
secondary star mass of 0.40~M$_{\odot}$ and a 
minimum mass ratio of 0.53, consistent to within 1$\sigma$ of the mass ratio found
from the full orbital solution for the system, 0.73$\pm$0.23.  The
approximate H-band flux ratio measured from the cross-correlation is 0.6$\pm$0.1. 
The K-band flux ratio found in the best PSF fit to the January, 2008
AO images is 0.66$\pm$0.18.
For a M2.5 $+$ K7 pair, the spectral types which provided the best correlation (\S 3), models of \citet{pal99} and \citet{bar98} imply an H-band flux ratio of 0.4$\pm$0.1 based on components with a primary T$_{eff}=$3900~K and a secondary T$_{eff}=$3400~K \citep{2003ApJ...593.1093L,joh66}.  Taken together, these properties suggest a relatively red color for the secondary star, hindering detection in visible light.  

The IR primary and secondary velocities, presented in the Wilson plot
(Figure 2), imply a range of mass ratios, from 0.73$-$0.83, well within
the range we obtain by combining the IR and visible light data in a full
orbital solution, $\sim$0.50$-$0.93.
Because the full solution includes phase information, it is generally
more reliable, although for RX~J0529.3$+$1210 the sparse
phase coverage during the epochs of largest velocity separation yields large
uncertainties in the orbital solution.  Clearly more data are merited.
For the case of a mass ratio relatively
close to unity in this system (i.e. 0.8$-$1.0), it is necessary to 
explain the lack of a visible light detection of the secondary star.
In the following paragraphs we outline several scenarios that would account
for this.

A relatively red secondary and yet a mass ratio close to unity is possible
if a third component is present, as first suggested by \citet{tor02}. If
the secondary star is actually a binary
pair, with a very small separation and similar component masses, then the
spectral types could be consistent with the best-fitting TODCOR templates.
This inner binary pair would necessarily
be in an orbit in a plane relatively perpendicular to our line of sight, such
that the large velocity changes would not be obvious. This scenario
would imply a near-unity mass ratio yet a smaller near-IR
flux ratio, and a red enough
secondary system to evade visible light detection.  The drawback is
that it requires a very specific geometry.   Such a companion might
also account for the 24~$\mu$m excess.

A heavily spotted secondary star could mimic a lower temperature source,
resulting in a best fitting secondary M2.5 template, a mass ratio close to 1.0, and a flux ratio $<$1 in the near-IR.  Again, the reddening effect of the cooler temperature in the spot covered areas would increase the difficulty of visible light detection.  For two T$_{eff}=$3900~K stars, if one
has 50\% of its surface covered with a 2900~K spot \citep{bou89}, the secondary/primary bolometric flux ratio would be $\sim$0.7.

For objects with ages between 10 and 30 Myr, some models \citep{bar98,pal99} of PMS
evolution indicate that targets in the mass range of the RX~J0529.3$+$1210
primary star, 0.6$-$0.8~M$_{\odot}$, are located near the transition between
the convective Hyashi track and the radiative Henyey track.  As a result,
models show a sharp turn in the mass tracks towards higher temperatures.
A similar but slightly lower mass secondary star that has not yet turned
this corner off of the Hyashi track, and indeed, objects with masses lower
than $\sim$0.5~M$_{\odot}$ never will, may have a similar mass as the
primary star but a much lower temperature.

These models for the system behavior are highly speculative.  The
absence of an unambiguous secondary star detection in visible light
suggests that the mass ratio is {\it not} close to unity.
A value in the 0.7 to 0.8 range is most likely based on data to date.

\subsection{Binary Formation Scenarios and the Potential for Planets}

The formation of such a highly eccentric multiple is challenging to
understand as a primordial event given current theories of star formation
\citep{2005fost.book.....S}.  Figure 4 shows that pre$-$main sequence spectroscopic
binaries are observed over a wide range of eccentricities.  Although RX~J0529.3$+$1210
represents the maximum of that range, it does not stand out particularly from the
eccentricity distribution.  Thus, some form of dynamical evolution may have taken place in
this system, but, if so, it is not distinguished by an outlying eccentricity.
Possibilities for dynamical evolution range from an improbable capture event to
disk excitation of stellar eccentricity \citep{mat92}. If indeed this system
is $>$10~Myr old, only fossil evidence, such
as the detected 24~$\mu$m excess, might remain of the primordial, presumably
massive disk(s) responsible.  In a variation of the interactions proposed
by \citet{rei00}, a three body dynamical encounter between an object
from outside of the binary and a companion short period binary could have
stimulated the eccentricity of the system and tightened the orbit of a
close (secondary) pair (\S 4.2).  The eventual formation of a circumbinary (or
circumtriple) debris disk could have been stimulated by dynamical
disruption of the disks in the system in any one of these scenarios.

The formation of planets in binary systems has been observationally supported
in recent years and is of primary significance since a majority of stars have
companions \citep{egg07}. \citet{cun07} have
determined the strict criteria for classifying stable and unstable planetary
orbits in binary systems, however, their analysis is restricted to circular
orbits. Many of the extra-solar planets observed in binaries are in wide
separation systems, where the mass ratio for the stellar components is near
unity and the eccentricity of the system is low.  \citet{qui07} have modeled the
formation of terrestrial planets around individual stars in binaries and
find that, for periastron
distances of $<$5~AU, such planet formation is restricted.  Considering that the
periastron distance in the RX~J0529.3$+$1210 system is only 0.15~AU, we
conclude that the probability of even a low-mass
circumstellar planet in RX~J0529.3$+$1210 is remote.  The trend determined
by \citet{qui06} towards the formation of fewer circumbinary
terrestrial planets in systems with apastron distances of $>$0.2~AU and
non-zero eccentricities also bodes for poor circumbinary planetary stability around
the stars in RX~J0529.3$+$1210.

\subsection{Improving our Understanding of the RX~J0529.3$+$1210 System: Future Work}

Clearly it is imperative to improve the orbital solution for RX~J0529.3$+$1210,
particularly the measurement of the mass ratio.  Visible light and especially
infrared data are critical to obtain, preferably at high precisions.
RX~J0529.3$+$1210 passes through its next maximum velocity separation in
late December, 2009.  Densely sampled
high signal-to-noise spectra taken during these epochs will yield a
greatly improved precision for the orbital solution.

If RX~J0529.3$+$1210 is located at a distance of only 90~pc, an apastron 
separation of 2.30~AU implies a maximum projected separation on the sky of
$\sim$0.03$''$.  Much of the orbit of this system lies within reach of the
diffraction limit of the 85~m baseline of the Keck Interferometer, 0.005$''$.
Although the K$-$band magnitude of RX~J0529.3$+$1210, 9.2, is relatively faint
for observations with this facility, planned improvements to the system
may eventually enable resolved observations of this unusual binary,
permitting the determination of individual component masses.  The 90~pc distance
may also be confirmed by measuring the parallax, which is slightly greater than 10 mas
at this distance, and is at the achievable limit of current instrumentation.

\section{Summary}

\citet{tor02} reported the orbital solution for RX~J0529.3$+$1210 based on single-lined
spectroscopic data; we have elaborated on this solution using near-IR data to identify the
spectrum of the secondary star and determine the mass ratio of the system.
The IR data alone presented in a primary vs. secondary velocity plot (Figure 2) indicate a
mass ratio of 0.78$\pm$0.05.  In concert with the visible light data, the full orbital
solution yields a more uncertain
mass ratio of 0.73$\pm$0.23 (Table 3).  Other orbital parameters are in good agreement
with the results of \citet{tor02}.  This system is the most eccentric pre$-$main
sequence double$-$lined binary known to date (Figure 4) and thus presents challenges to
observation.  It is an unlikely site for the formation of circumstellar
planets, or of circumbinary planets with orbits within several AU of the stars.

The fact that the secondary component was not identified in visible light
suggests a relatively faint and/or red companion.  For an estimated primary mass
of 0.75~M$_{\odot}$ the minimum mass of the secondary is 0.40~M$_{\odot}$.  This
implies a minimum mass ratio of 0.53, consistent with the value in Table 3.  A robust
determination of the mass ratio along with the best estimate for the primary star
mass will eventually provide an improved estimate
for the secondary mass.

On the basis of the center-of-mass velocity, common proper motion, a lack of near-IR
excess, a tentative K-band AO detection of the secondary, lithium equivalent width,
and 24~$\mu$m {\it Spitzer} excess evidence for a debris disk, we argue that the
RX~J0529.3$+$1210 system is a member of the 32 Ori moving group, recently
identified by \citet{mam07}.  Assuming a distance of
90~pc, we estimate an age for RX~J0529.3$+$1210 of 15$\pm$5~Myr, consistent with that
of \citet{mam07} for 32 Ori.

It is imperative to improve the orbital elements of RX~J0529.3$+$1210 in
order to determine a precise mass ratio for the system.  In conjunction
with future interferometric observations, this will ultimately yield
component masses for this system.  Given its unusual eccentricity,
proximity, young age, and debris disk, such observations are likely to reveal
clues to the formation of this unique binary system as well as additional data
for the improvement of pre-main sequence models.

\begin{acknowledgments}
The authors are grateful to K. Kilts for conducting the preliminary
reduction of these data, to T. Barman and C. Johns-Krull for helpful discussions,
and to G. Torres for catching an error in our original manuscript.
We thank M. Morales Calderon, D. Barrado y Navascu\'es, and J. Stauffer for
sharing the results of their {\it Spitzer} MIPS data in advance of publication.
We appreciate the input of E. Mamajek, on the traits of the 32 Ori moving group,
and we are grateful to the anonymous referee for detailed and thoughtful
recommendations which improved this paper.  This research was funded by
NSF grant AST 04-44017 (to LP) and the associated REU supplement, and by
NASA Space Grant, through Northern Arizona University, which has provided
support for GNM, as well as NSF grant AST 06-07612 (to MS).
This work made use of the SIMBAD reference database, the NASA
Astrophysics Data System, and the data products from the Two Micron All
Sky Survey, which is a joint project of the University of Massachusetts
and the Infrared Processing and Analysis Center/California Institute
of Technology, funded by the National Aeronautics and Space
Administration and the National Science Foundation.
Data presented herein were obtained at the W.M. Keck
Observatory from telescope time allocated to the National Aeronautics
and Space Administration through the agency's scientific partnership
with the California Institute of Technology and the University of
California. The Observatory was made possible by the generous
financial support of the W.M. Keck Foundation.
We recognize and acknowledge the
significant cultural role that the summit of Mauna Kea
plays within the indigenous Hawaiian community and are
grateful for the opportunity to conduct observations
from this special mountain.
\end{acknowledgments}

\clearpage

\pagestyle{empty}

\begin{deluxetable}{lc}
\tablewidth{0pt}
\tablecaption{Properties of RX~J0529.3$+$1210
\label{tbl-1}}
\tablehead{}
\startdata
R.A.(J2000) = 05:29:18.8\\
Dec.(J2000) = +12:09:30\\
V (mag) = 12.86\\
J (mag) = 10.05 $\pm$ 0.020\\
H (mag) = 9.40 $\pm$ 0.023\\
K (mag) = 9.19 $\pm$ 0.020\\
SpT\tablenotemark{a} = K7-M0\\
H$\alpha$ EW\tablenotemark{b} (\AA) = $-$2.0\\
Li EW (\AA) = 0.27\tablenotemark{a}, 0.35\tablenotemark{b}\\
v$_A$sin$i$\tablenotemark{a} $=  $18 $\pm$ 3 km s$^{-1}$\\
\enddata

\tablenotetext{a}{Torres et al. (2002)}
\tablenotetext{b}{Magazz\`{u} et al. (1997)}

\end{deluxetable}

\clearpage

\pagestyle{empty}

\begin{deluxetable}{lccrrl}
\tablewidth{0pt}
\tablecaption{Summary of Observations and Analysis\label{tbl-2}}
\tablehead{
\colhead{UT Date of} & \colhead{$~$} &
\colhead{$v_1$}  & \colhead{$v_2$  }  & \colhead{$~$} \\
\colhead{Observations} & \colhead{Heliocentric Julian Day} &
\colhead{(km s$^{-1}$)} & \colhead{(km s$^{-1}$)} & \colhead{Phase}}
\startdata
2002 Jan 1 & 2452275.88  & 24.7 & 12.3 & 0.183 \\
2002 Feb 5 & 2452310.87  & 26.9 & 13.9 & 0.258 \\
2002 Dec 14 & 2452622.97  & 19.5 & 14.9 & 0.933 \\
2004 Jan 28 & 2453032.76  & 15.4 & 19.7 & 0.818 \\
2004 Dec 26 & 2453365.99  & 2.3 & 42.3 & 0.538 \\
2008 Sep 15 & 2454725.15  & 15.3 & 17.9 & 0.498 \\
\enddata

\end{deluxetable}

\clearpage

\pagestyle{empty}

\begin{deluxetable}{lc}
\tablewidth{0pt}
\tablecaption{Orbital Elements and Derived Properties  \label{tbl-3}}
\tablehead{}
\startdata
$P = 461.89 \pm 0.15$ days\\
$\gamma = 18.38 \pm 0.30$ km s$^{-1}$\\
$K_1 = 22.76 \pm 1.59 $ km s$^{-1}$\\
$K_2 = 31.25 \pm 9.44 $ km s$^{-1}$\\
$e = 0.88 \pm 0.02$ \\
$\omega = 108 \pm 4$ degrees\\
$T = 2455187.1 \pm 0.46$ MJD\\
$M_1$ sin$^3 i = 0.454 \pm 0.312$  \\
$M_2$ sin$^3 i = 0.330 \pm 0.317$ \\
$q = M_2/M_1 = 0.73 \pm 0.23 $ \\
$a_1$ sin $i = (67.98 \pm 2.05) \times 10^6$ km\\
$a_2$ sin $i = (93.32 \pm 27.57) \times 10^6$ km\\
\enddata
\end{deluxetable}

\clearpage

\begin{figure*}
\includegraphics[angle=0,width=5.5in]{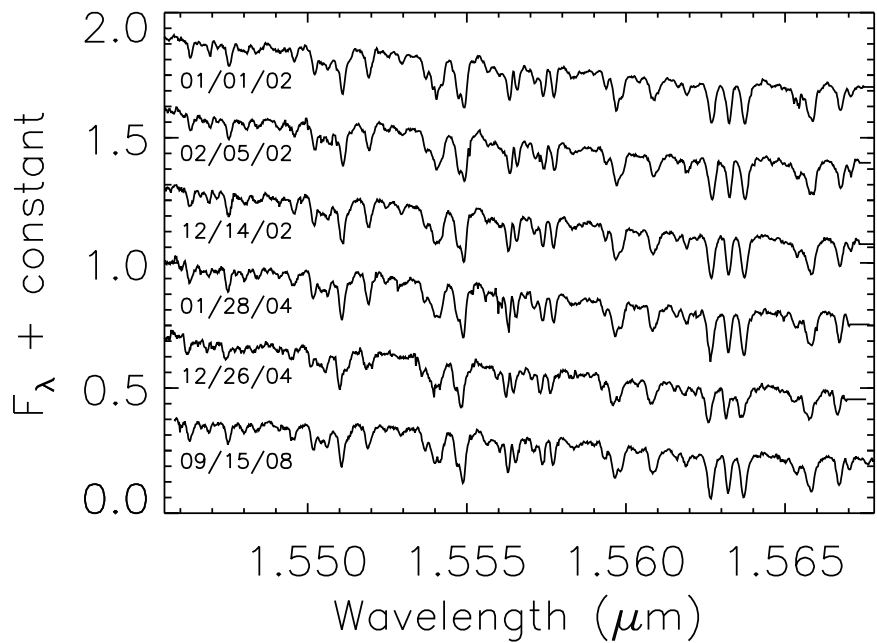}
\caption{Five epochs of spectra for RX~J0529.3$+$1210, with heliocentric
  corrections applied; UT dates of the observations are indicated.  Note the similarity between observations resulting from small radial velocity separations.}
\label{fig:rxj05293}
\end{figure*}

\begin{figure*}
\includegraphics[angle=0,width=5.5in]{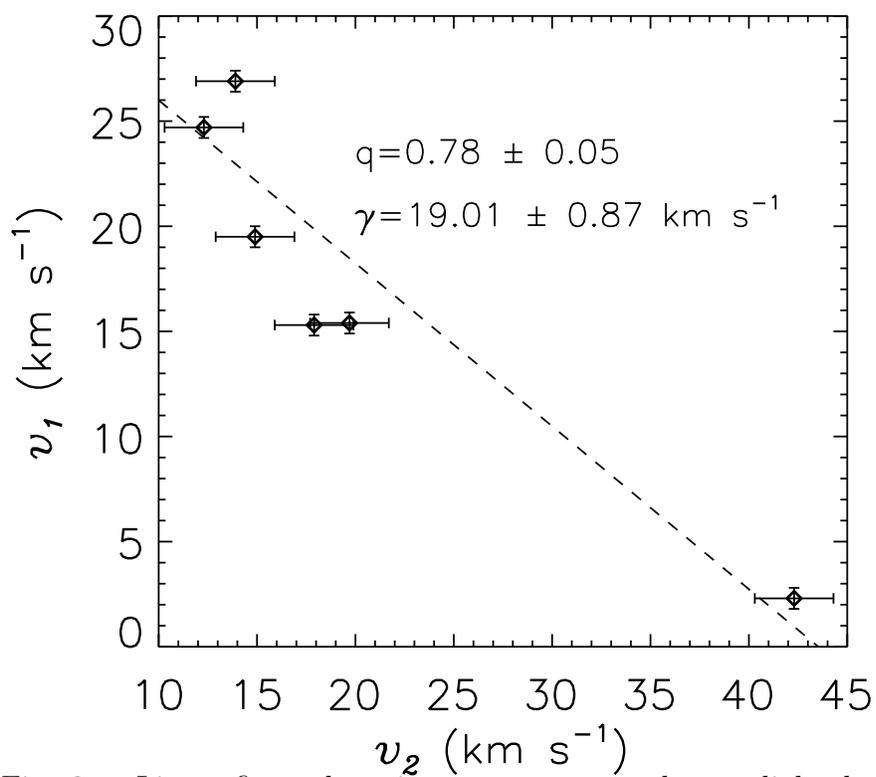}
\caption{Linear fit to the primary versus secondary
radial velocities for RX~J0529.3$+$1210 following \citet{wil41}.  The mass ratio, q, is the negative of the slope of the fit and the center-of-mass velocity is determined by the equation $\gamma$=(y-intercept)/(1$+$q).}
\label{fig:W41_rxj05293}
\end{figure*}

\begin{figure*}
\includegraphics[angle=0,width=6.0in]{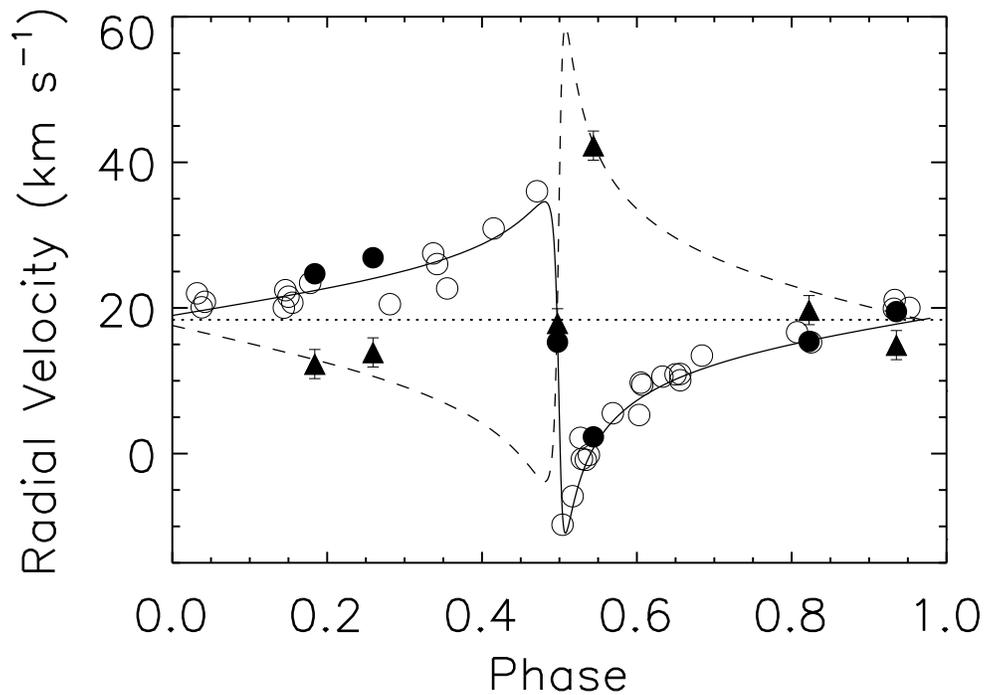}
\caption{Radial velocity as a function of phase for RX~J0529.3$+$1210.  
The circles represent the primary star data, and the triangles secondary star
data. Filled symbols are data from our observations, and hollow symbols are data
from Torres et al. (2002). The full orbital solution is represented with a solid line
for the primary star and a dashed line for the secondary star. The dotted horizontal
line indicates the system's center-of-mass velocity.  Uncertainties in the primary star
RVs are 0.5~km~s$^{-1}$, smaller than the symbols used.  The secondary
star RV uncertainties are 2.0~km~s$^{-1}$, as shown.} 
\label{fig:tprv}
\end{figure*}

\begin{figure*}
\includegraphics[angle=0,width=6.0in]{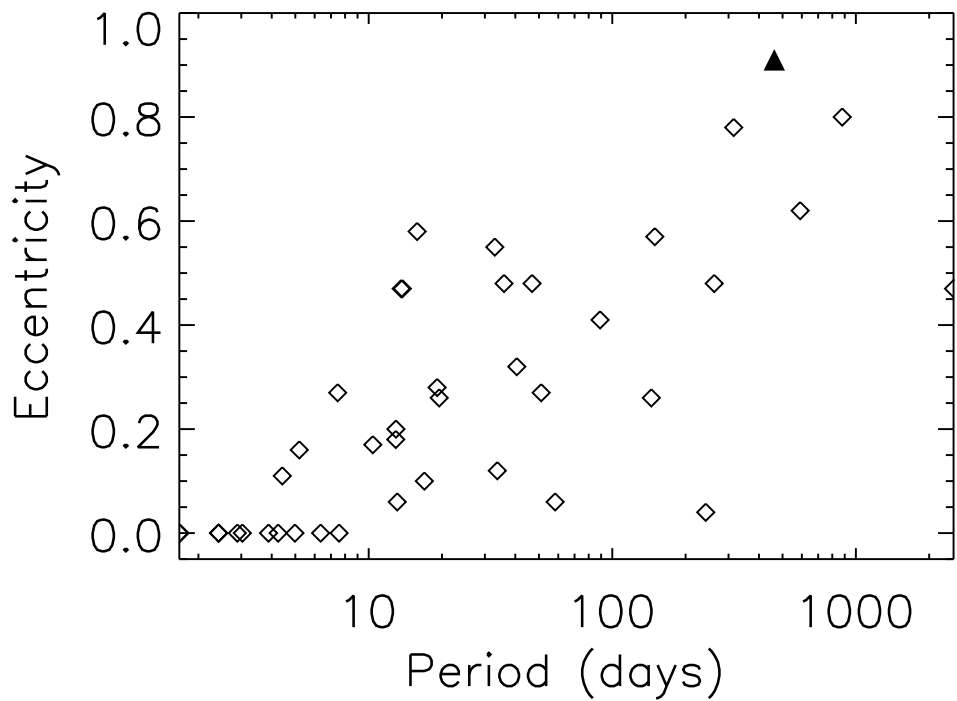}
\caption{Eccentricities of pre-main sequence spectroscopic binaries
as a function of period.  RX~J0529.3$+$1210 is indicated by the filled triangle.  Data from Table 3 of Melo et al. (2001) are indicated by diamonds.} 
\label{fig:melo}
\end{figure*}


\clearpage
\begin{thebibliography}{} 

\bibitem[Barrado y Navascu{\'e}s(2005)]{bar05} Barrado Y 
Navascu{\'e}s, D.\ 2005, Revista Mexicana de Astronomia y Astrofisica 
Conference Series, 24, 217

\bibitem[Barrado y Navascu{\'e}s et al.(2007)]{bar07} Barrado 
y Navascu{\'e}s, D., et al.\ 2007, \apj, 664, 481

\bibitem[Baraffe et al.(1998)]{bar98} Baraffe, I., Chabrier, G.,
Allard, F., \& Hauschildt, P. H. 1998, \aap, 337, 403

\bibitem[Beck et al.(2003)]{2003ApJ...583..358B} Beck, T.~L., Simon, M., 
\& Close, L.~M.\ 2003, \apj, 583, 358 

\bibitem[Boden et al.(2005)]{2005ApJ...635..442B} Boden, A.~F., et al.\ 
2005, \apj, 635, 442 

\bibitem[Bouvier \& Bertout(1989)]{bou89} Bouvier, J., \& Bertout, C.\ 1989, \aap,  
211, 99

\bibitem[Broeg et al.(2006)]{bro06} Broeg, C., Joergens, V., Fern\'{a}ndez,
  M., Husar, D., Hearty, T., Ammler, M., \& Neuhauser, R. 2006, \aap, 450, 1135

\bibitem[Cuntz et al. (2007)]{cun07} Cuntz, M., Eberle, J., \& Musielak,
  Z.E. 2007, \apj, 669, L105

\bibitem[Dolan \& Mathieu(2001)]{dol01} Dolan, C.~J., \& Mathieu, R.~D.\ 2001, \aj, 121, 2124


\bibitem[Eggenberger \& Udry (2007)]{egg07} Eggenberger, A., \& Udry, S. 2007, in
  Planets in Binary Star Systems, ed. N. Haghighipour (New York: Springer), in
  press (astro-ph/0705.3173)
  

\bibitem[Hauschildt et al.(1999)]{1999ApJ...512..377H} Hauschildt, P.~H., 
Allard, F., \& Baron, E.\ 1999, \apj, 512, 377 

\bibitem[Herbig(1978)]{her78} Herbig, G.~H.\ 1978, in Problems 
of Physics and Evolution of the Universe, ed. L.V. Mirzoyan (Yerevan: Armenian Acad. Sci.), 171 

\bibitem[Johnson(1966)]{joh66} Johnson, H.~L.\ 1966, \araa, 4, 193 

\bibitem[Lada(2006)]{2006ApJ...640L..63L} Lada, C.~J.\ 2006, \apjl, 640, 
L63 

\bibitem[Luhman et al.(2003)]{2003ApJ...593.1093L} Luhman, K.~L., Stauffer, 
J.~R., Muench, A.~A., Rieke, G.~H., Lada, E.~A., Bouvier, J., 
\& Lada, C.~J.\ 2003, \apj, 593, 1093 

\bibitem[Magazzu{\`u} et al.(1997)]{mag97} Magazz{\`u}, A., Mart\'{i}n, E. L., Sterzik,
  M. F., Neuhauser, R., Covino, E., \& Alcala, J.M. 1997, \aap, 124, 449

\bibitem[Magazz{\`u} et al.(1999)]{mag99} Magazz{\`u}, A., Umana, G.,
\& Mart{\'{\i}}n, E.~L.\ 1999, \aap, 346, 878 

\bibitem[Mamajek(2007)]{mam07} Mamajek, E.~E.\ 2007, IAU 
Symposium, 237, 442

\bibitem[Mart{\'{\i}}n(1998)]{1998AJ....115..351M} Mart{\'{\i}}n, E.~L.\ 
1998, \aj, 115, 351 

\bibitem[Mathieu(1992)]{mat92} Mathieu, R.~D.\ 1992, in Binaries 
as Tracers of Stellar Formation.~Proceedings of a Workshop held in 
Bettmeralp, Switzerland, Sept.~1991. (Cambridge: Cambridge Univ. Press), 155

\bibitem[Mathieu et al.(2000)]{2000prpl.conf..703M} Mathieu, R.~D., Ghez, 
A.~M., Jensen, E.~L.~N., \& Simon, M.\ 2000, Protostars and Planets IV, ed. V. Mannings, A.P. Boss \& S.S. Russell (Tucson: Univ. Arizona Press), 703

\bibitem[Mazeh et al.(2002)]{maz02} Mazeh, T., Prato, L., Simon, M.,
Goldberg, E., Norman, D., \& Zucker, S. 2002, \apj, 564, 1007

\bibitem[Mazeh et al.(2003)]{2003ApJ...599.1344M} Mazeh, T., Simon, M., 
Prato, L., Markus, B., \& Zucker, S.\ 2003, \apj, 599, 1344 

\bibitem[McLean et al.(1998)]{mcl98} McLean, I. S., et al. 1998,
SPIE, 3354, 566

\bibitem[McLean et al.(2000)]{mcl00} McLean, I. S., Graham, J. R.,
Becklin, E. E., Figer, D. F., Larkin, J. E., Levenson, N. A., \& Teplitz,
H. I. 2000, SPIE, 4008, 1048

\bibitem[Melo et al. (2001)]{mel01} Melo, C.H.F., Covino, E., Alcal\'{a},
  J.M., \& Torres, G. 2001, \aap, 378, 898

\bibitem[Neuhauser et al.(1995)]{neu95} Neuhauser, R., Sterzik, M.~F.,
Torres, G., \& Martin, E.~L.\ 1995, \aap, 299, L13 

\bibitem[Neuhauser et al.(1997)]{neu97} Neuhauser, R., Torres, G.,
Sterzik, M.~F., \& Randich, S.\ 1997, \aap, 325, 647 

\bibitem[O'Dell(1998)]{1998AJ....115..263O} O'Dell, C.~R.\ 1998, \aj, 115, 
263 

\bibitem[Palla \& Stahler(1999)]{pal99} Palla, F., \& Stahler, S.~W.\ 1999, \apj, 525, 772

\bibitem[Palla \& Stahler(2001)]{2001ApJ...553..299P} Palla, F., \& Stahler, S.~W.\ 2001, \apj, 553, 299

\bibitem[Perryman et al.(1997)]{1997A&A...323L..49P} Perryman, M.~A.~C., et al.\ 1997, \aap, 323, L49 

\bibitem[Prato et al.(2002a)]{pra02a} Prato, L., Simon, M., Mazeh, T., Zucker, S., \&  McLean, I. S. 2002a, \apjl, 579, L99

\bibitem[Prato et al.(2002b)]{pra02b} Prato, L., Simon, M.,
Mazeh, T., McLean, I. S., Norman, D., \& Zucker, S. 2002b, \apj, 569, 863


\bibitem[Press et al.(1992)]{1992nrfa.book.....P} Press, W.~H., Teukolsky, 
S.~A., Vetterling, W.~T., \& Flannery, B.~P.\ 1992, Numerical Recipes in Fortran: The Art of Scientific Computing, (2nd edn.; Cambridge: Cambridge Univ. Press)

\bibitem[Quintana \& Lissauer(2006)]{qui06} Quintana, E.~V.,
\& Lissauer, J.~J.\ 2006, Icarus, 185, 1

\bibitem[Quintana et al.(2007)]{qui07} Quintana, E.~V., 
Adams, F.~C., Lissauer, J.~J., \& Chambers, J.~E.\ 2007, \apj, 660, 807

\bibitem[Reipurth(2000)]{rei00} Reipurth, B.\ 2000, \aj, 120, 
3177

\bibitem[Schaefer et al.(2008)]{2008AJ....135.1659S} Schaefer, G.~H., Simon, M., Prato, L., \& Barman, T.\ 2008, \aj, 135, 1659 

\bibitem[Siess et  al.(2000)]{2000A&A...358..593S} Siess, L., Dufour, E., \& Forestini, M.\ 2000, \aap, 358, 593 

\bibitem[Stahler \& Palla(2005)]{2005fost.book.....S} Stahler, S.~W., \&
  Palla, F.\ 2005, The Formation of Stars.~Wiley-VCH, Weinheim
  
\bibitem[Stassun et al.(2007)]{2007ApJ...664.1154S} Stassun, K.~G., 
Mathieu, R.~D., \& Valenti, J.~A.\ 2007, \apj, 664, 1154 

\bibitem[Steffen et al.(2001)]{2001AJ....122..997S} Steffen, A.~T., et al.\ 
2001, \aj, 122, 997 

  
\bibitem[Strom et al.(1989)]{1989AJ.....98.1444S} Strom, K.~M., Wilkin, 
F.~P., Strom, S.~E., \& Seaman, R.~L.\ 1989, \aj, 98, 1444 

\bibitem[Torres et al.(2002)]{tor02} Torres, G., Neuhauser, R., \& Guenther,
  E. W. 2002, \aj, 123, 1701

\bibitem[Trilling et al.(2008)]{tri08} Trilling, D.~E., et 
al.\ 2008, \apj, 674, 1086

\bibitem[Wilson(1941)]{wil41} Wilson, O. C. 1941, \apj, 93, 29

\bibitem[Wizinowich et al.(2000)]{2000SPIE.4007....2W} Wizinowich, P.~L., Acton, D.~S., Lai, O., Gathright, J., Lupton, W., \& Stomski, P.~J.\ 2000, \procspie, 4007, 2 

\bibitem[Zacharias et al.(2004)]{2004AJ....127.3043Z} Zacharias, N., Urban, 
S.~E., Zacharias, M.~I., Wycoff, G.~L., Hall, D.~M., Monet, D.~G., 
\& Rafferty, T.~J.\ 2004, \aj, 127, 3043 

\bibitem[Zucker \& Mazeh(1994)]{zuc94} Zucker, S.,
\& Mazeh, T. 1994, \apj, 420, 806

\end{thebibliography}
\end{document}